\documentstyle[prl,aps,psfig]{revtex}

\begin{document} 
\draft

\title{Footprints in Sand: The Response of a Granular Material to
       Local Perturbations}

\author{Junfei Geng, D. Howell, E. Longhi, and R.~P. Behringer}

\address{Center for Nonlinear and Complex Systems, Duke University,
Durham NC, 27708-0305, USA} 

\author{G. Reydellet, L. Vanel and E. Cl\'{e}ment}

\address{Universit\'{e} Pierre et Marie Curie, Paris 75231, France}

\author{S. Luding}

\address{ICA1, University of Stuttgart, Pfaffenwaldring 27, 
               70569 Stuttgart, Germany}

\date{\today} 

\maketitle

\begin{abstract} 
We experimentally determine ensemble-averaged responses of granular
packings to point forces, and we compare these results to recent
models for force propagation in a granular material.  We used 2D
granular arrays consisting of photoelastic particles: either disks or
pentagons, thus spanning the range from ordered to disordered
packings.  A key finding is that spatial ordering of the particles is
a key factor in the force response.  Ordered packings have a
propagative component that does not occur in disordered packings.
\end{abstract}


Granular systems have captured much recent interest due to their rich
phenomenology, and important applications \cite{reviews}.  Even in the
absence of strong spatial disorder of the grains, static arrays show
inhomogeneous spatial stress profiles called stress (or force)
chains\cite{Roux97}.  Forces are carried primarily by a tenuous
network that is a fraction of the total number of grains.

A fundamental unresolved issue concerns how granular materials respond
to applied forces, and there are several substantially different
models.  A broad group of conventional continuum models (e.g.\
elasto-plastic, $\ldots$) posit an elastic response for material up to
the point of plastic deformation \cite{continuum}.  The stresses in
portions of such a system below plastic yield have an elastic response
and satisfy an elliptic partial differential equation (PDE); those
parts that are plastically deforming satisfy a hyperbolic PDE.
Several fundamentally different models have recently been
proposed. The $q$-model of Coppersmith et al.\ \cite{coppersmith}
assumes a regular lattice of grains, and randomness is introduced at
the contacts.  This model successfully predicts the distribution of
forces in the large force limit, as verified by several static and
quasistatic experiments and models
\cite{coppersmith,mueth_98,howell_99}.  In the continuum limit, this
model reduces to the diffusion equation, since the forces effectively
propagate by a random walk.  Another model (the Oriented Stress
Linearity--OSL--model) of Bouchaud et al.\cite{bouchaud}, has a
constitutive law, justified through a microscopic model, of the form
$\sigma_{zz} = \mu \sigma_{xz} + \eta \sigma_{xx}$ (in 2D) in order to
close the stress balance conditions $\partial \sigma_{ij}/\partial
x_{j} = \rho g_i$.  This leads to wave-like hyperbolic PDEs describing
the spatial variation of stresses.  In later work, these authors
considered weak randomness in the lattice \cite{claudin}, and proposed
a convection-diffusion (C-D) equation.  A model proposed by Kenkre et
al.\cite{kenkre} combines pure wave propagation and diffusion in the
telegraph equation.  Recently, Luding et al.\ \cite{luding_model} and
Bouchaud et al. \cite{claudin_model} have proposed models that account
for force fluctuations on both the length scale of grains and on the
length scale of stress chains.

The range of predictions among the models is perhaps best appreciated
by noting that the different pictures predict qualitatively different
PDEs for the variation of stresses within a sample: e.g.\ for
elasto-plastic models an elliptic or hyperbolic PDE; for the
$q$-model, a parabolic PDE; and for the OSL model without randomness,
a hyperbolic PDE.  The impact of equation type extends to the boundary
conditions needed to determine a solution: e.g.\ hyperbolic equations
require less boundary information than an elliptic equation.

\begin{figure}[htb]
\center{\parbox{3.375in}{\psfig{file=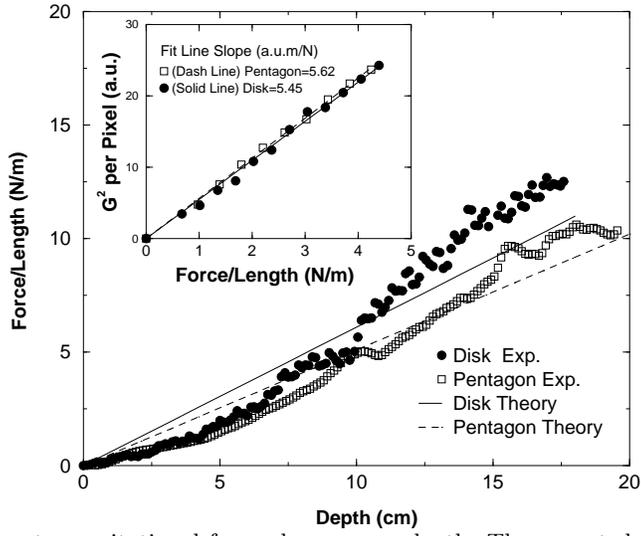,width=3.375in}}}
\caption{Hydrostatic pressure due to gravitational force alone versus
depth. The expected slopes of the stress-height curves are calculated
from the known packing fraction ($\gamma=0.91$ for disks; $\gamma
\simeq 0.75$ for pentagons). Inset shows the multi-particle $G^{2}$
calibration by applying known loads to the upper surface of the
layer.}
\label{fig:misc}
\end{figure}

Here, we explore these issues through experiments on a 2D granular
system consisting of photoelastic (i.e birefringent under strain)
polymer particles \cite{howell_99} that are either disks or pentagons.
By viewing the particles through an arrangement of circular polarizers
(a polariscope) it is possible to characterize the stress on the
particles \cite{howell_99}.  In the images shown below, bright regions
correspond to locally large force/stress.  Near a contact, the
stresses within a particle are very nonuniform.  This leads to a
series of light and dark bands in the polariscope image intensity,
$I$, with the density of bands increasing monotonically with the force
at the contact.  We exploit this fact to produce a force calibration
in terms of $G^2 \equiv |\nabla I|^2$, since the mean $\langle G^2
\rangle$ over the scale of a disk grows with the contact forces.  We
obtained this calibration either by applying known forces $F$ to the
boundary of a small number of particles and at the same time measuring
$\langle G^2 \rangle$, or by applying uniform loads to the upper
surface of a large rectangular sample.  As a confirmation that this
technique measures the force (pressure), we show the hydrostatic
pressure vs. depth for two different samples in Fig.~\ref{fig:misc}.

We used two different particle types, disks and pentagons, in order to
construct packings of varying degrees of spatial order/disorder.  For
monodisperse disks, we obtained a highly ordered packing.  With
bidisperse distributions, we modified the disorder in a controlled way
\cite{luding} characterized by the parameter ${\mathcal{A}} = \langle
a \rangle ^2/ \langle a^2 \rangle$.  The brackets refer to averages
over the sample for powers of the disk diameters $a$.  To vary
${\mathcal{A}}$, we used mixtures of two disk diameters: $a_1 \approx
0.7\,{\rm cm}$ and $a_2 \approx 0.9\,{\rm cm}$.  Both types of disks
had a thickness of 0.64~cm.  The pentagons had the same thickness as
the disks, and a side length of 0.7~cm.  A sample, typically $\sim 60$
particles wide and $25$ particles high, was placed in a nearly
vertical plane.  To keep the sample from collapsing, it rested against
a smooth powder-lubricated glass plate that was inclined from the
vertical by a small angle, $\stackrel{<}{\sim} 3^o$; hence minimizing
the friction with the plate.  The sample rested on a rigid metal base,
and was confined at the sides by rigid metal bars.  The samples were
prepared by gently adding particles to the upper surface until the
full amount of particles was in place.

\begin{figure}[htb]
\center{\parbox{3.375in}{\psfig{file=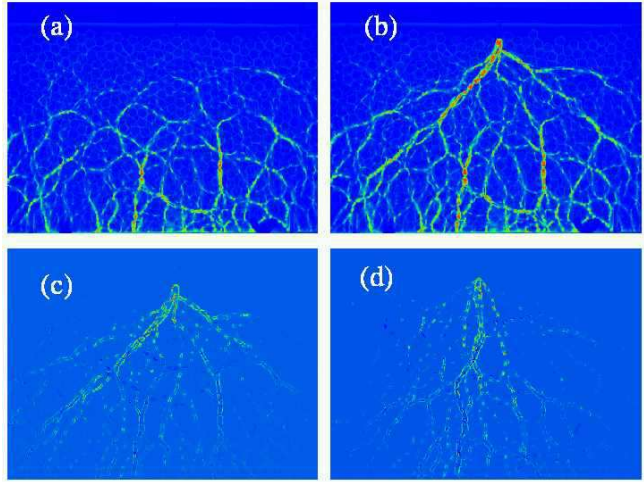,width=3.375in}}}
\caption{Images for pentagons showing a) the force pattern due to
hydrostatic head. b) the combined response to gravity and a point
source. c) the response to a point force--after subtracting image a
from image b. d) similar to c, but with a different grain
configuration.}
\label{fig:images}
\end{figure}

Fig.~\ref{fig:images}a shows a typical polariscope image in the
absence of any applied force, besides gravity.  The stress at the
bottom of the sample is larger than at the top due to gravitational
head, and the mean stress was a linearly increasing function of depth.

We measured the system response by placing a known weight carefully on
top of one particle in the approximate center of the sample, producing
a local vertical force (Fig.~\ref{fig:images}b).  We then removed the
weight.  For the monodisperse and pentagonal packings described below,
we used only responses where the particles remained undisturbed by
this process.  For polydisperse disk samples, some small rearrangement
of the surface grains always occured, however.  For each realization,
we obtained the stress difference between successive images
(Fig.~\ref{fig:images}c), containing only the response from the point
perturbation, with the linear hydrostatic head effects removed.

We repeated this process for many different rearrangements of the
particles, typically 50 times for each set of particles.  The images
of Fig.~\ref{fig:images}c,d show that the responses are complicated
and differ significantly from realization to realization.  There are
at least two approaches to address the large variability among
realizations: 1) obtain information on an ensemble of realizations; or
2) perturb the system over a larger number of grains in order to
obtain a larger-scale averaging.  Here, we pursue the first option.

\begin{figure}[htb]
\center{\parbox{3.375in}{\psfig{file=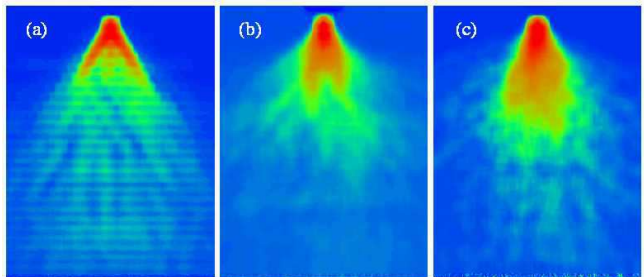,width=3.375in}}}
\caption{Mean response for 50 trials of a 50 g point force for (a) a
uniform hexagonal packing of disks, (b) a bimodal packing disks with
${\mathcal{A}}=0.99$, and (c) for pentagons. The size of each image is
about $18.0 \times 13.5$~cm$^2$.}
\label{fig:means}
\end{figure}

\begin{figure}[htb]
\center{\parbox{3.375in}{\psfig{file=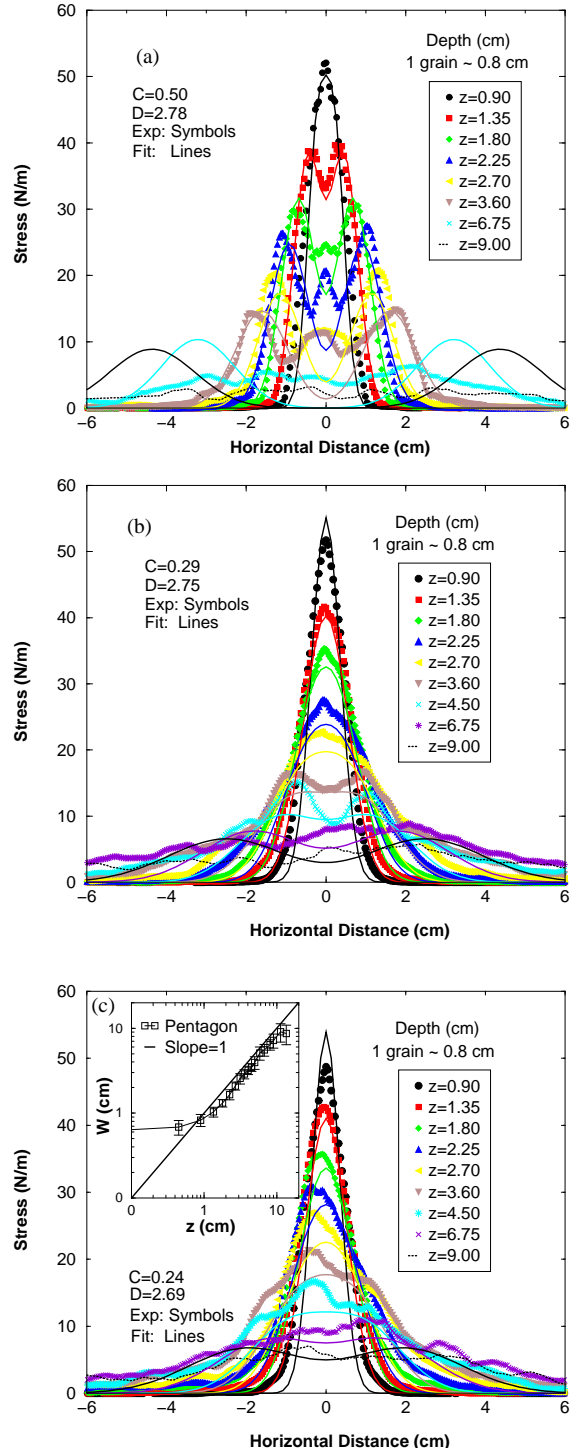,width=3.2in}}}
\caption{Photoelastic response (force per length--stress in 2D) to a
point force, vs.\ horizontal distance, $x$, at various depths, $z$,
from the source. (a) for ordered disks, (b) for bimodal disks, and (c)
for pentagons.  Also shown are fits of the response to the
convection-diffusion model. In all fits, $c$ is dimensionless and $D$
is in units of pixels (1 pixel $=$ 0.45 mm).  Inset part c: $W$
vs. $z$ for the response for pentagons.}
\label{fig:quantitative}
\end{figure}

In Fig.~\ref{fig:means}, we show using greyscale the average response
for monodisperse disks, a mixture of disks, and pentagons, for a point
load of $50\,{\rm g}$.  In Fig.~\ref{fig:quantitative}, we show more
quantitative data for the force response along a series of horizontal
lines at depths, $z$ (measured from the top).  These data show clear
differences depending on the packing geometry.  The monodisperse
packing shows two peaks emerging from the central source, broadening
with $z$, but remaining clearly identifiable.  For the bidisperse
packing, the peaks are also present, but much less sharply resolved.
Data for the pentagons show no evidence for wave-like stress
propagation.  There is some noisy structure for the pentagon response
at the greatest depths.  We believe this is due to the fact that it is
relatively difficult to rearrange the interlocked pentagonal packing
at large depths; hence chains tended to appear there in nearly the
same places.  In any event, this effect only occurs deep in the pile.

A key question is then which model best describes these results.  The
answer clearly depends on the amount of spatial order in the packing.
For the ordered (monodisperse) packing, the C-D model proposed by
Claudin et al.\ \cite{claudin} provides reasonable agreement with the
data, as shown by the fit lines of Fig.~\ref{fig:quantitative}.
However, there are clear departures between the data and fits to the
C-D model, since the central peak persists to depths for which the C-D
model predicts no such feature, and the two side structures fall to
the noise level for $h > 8d$.  The recent models proposed by Luding et
al.\ \cite{luding_model} and by Bouchaud et al.\cite{claudin_model}
predict the presence of the central feature, and thus offer improved
fits (details to be discussed elsewhere).  For the disordered
packings, best characterized by pentagons, the one central peak might
be described by either diffusion or by elasticity.

It is possible to distinguish which of these is best from the width,
$W(z)$, of the response peak.  For a diffusive model, $W$ should grow
as $\sqrt{z}$.  For an elastic layer, the response to a point force is
a modified Lorentzian with $W \propto z$ for moderate $z$.  We show in
the inset of Fig.~\ref{fig:quantitative}c data (on log scales)
indicating that $W \propto z$ over much of the experimental range, as
in an elastic response.

The pentagon results, including the linear dependence of $W$ on $z$,
are consistent with recent experimental work by Reydellet et al.\
\cite{reydellet} for 3D disordered packings, with numerical
simulations by Moreau\cite{moreau}, and with the recent theoretical
work of Bouchaud et al.\cite{claudin_model}, who find a crossover to
an elliptic response beyond a length scale controlled by the level of
disorder.  Recent experiments have also been carried out by
Rajchenbach \cite{rajchenbach_00}, who used cuboidal blocks of
photoelastic material in a similar arrangement to that used here.
These experiments showed a parabolic outer envelope of the response.
Since it is difficult to know the nature of the intergrain contacts
and the extent of their order/disorder, it is also difficult to place
these measurements in the present context.

To conclude, we have measured the force response of 2D granular
packings to local force perturbations.  There are large variations for
any given response, and we consider averages over many realizations.
For packings with strong spatial order, the mean response has a strong
propagative part, and convection-diffusion or more recent models
\cite{luding_model,claudin_model} give a reasonable description.  As
the amount of disorder increases, the propagative component
diminishes, and the response is most like that of an elastic material.
Interestingly, the connection between order/disorder and the force
response may provide a way to characterize the disorder in granular
samples, something that may be important for understanding experiments
where the amount of disorder of a packing was clearly important
\cite{baxter_89}.  These results raise additional questions for future
work such as the need to investigate the vector character of force
propagation (e.g. forces applied at arbitrary angles to a surface,
forces applied in the interior).  Another important issue concerns the
statistical variability that can be expected in a single realization.

{\bf Acknowledgments} We appreciate helpful interactions with
P. Claudin, I. Goldhirsch, D. Schaeffer, and J.  Socolar.  The work of
SL was supported by the Deutsche Forschungsgemeinschaft (DFG) and the
National Science Foundation.  The work of GR, LV, and EC was supported
by PICS-CNRS $\#563$.  The work of JG, DH, EL, and RPB was supported
by the US National Science Foundation under Grant DMR-9802602, and
DMS-9803305, and by NASA under Grant NAG3-2372.

\end{document}